\documentclass[runningheads]{llncs}
\usepackage[T1]{fontenc}
\usepackage{amsmath}  % if amsmath is not already loaded
\usepackage{amssymb}
\usepackage{microtype}
\usepackage{subfigure}
\usepackage{booktabs} % for professional tables
\usepackage{graphicx}

\usepackage{mathtools}

\usepackage{xspace}

\newcommand{\AB}{\textit{AB}\xspace}

\newcommand{\BB}{\textit{BB}\xspace}
\newcommand{\micron}{{\textmu}m }
\newcommand{\ppm}{\,$\pm$\,}
\begin{document}
\title{Unsupervised Domain Adaptation for Cerebellum Segmentation}
% \author{Anonymous Authors}

\author{Xuan Li, Paule-J Toussaint, Alan Evans, Xue Liu}
\institute{McGill University}
% \institute{Anonymous University}
%
\maketitle              % typeset the header of the contribution
\begin{abstract}
Unsupervised domain adaptation (UDA) is a popular approach in medical imaging for transferring annotation knowledge from a labelled source domain to an unlabelled target domain. However, manual labelling of medical images is often time-consuming, particularly for high-resolution images with sparse annotations such as the cerebellar atlas with its convoluted foliations. The lack of comprehensive annotation has hindered studies of cerebellar involvement in normal brain function and disease.
Empirically, we find that distinct visual discrepancy due to different staining and spacing between sections, only single subject in each domain, and few available sections in the source domain create uncertainty and inconsistency when generating pseudo labels for the target dataset. These factors can prevent existing approaches from providing meaningful segmentation annotations. 
To address these challenges, we propose a Bayesian learning framework that imposes constraints on the pseudo labels and guides the segmentation model to learn from regions with the most confidence. We design a variance map to rectify the uncertainty using variational sampling. Our approach generates expert-approved semantic annotations, as validated by qualitative results, and achieves over 2\% loss reduction and over 3.7\% intersection-over-union (IoU) improvement compared to other approaches in quantitative experiments.

\end{abstract}

\section{Introduction}
\label{introduction}
\begin{figure}[ht!]
\centering
\setlength{\tabcolsep}{9pt} 
\begin{tabular}{cc}
\includegraphics[width=.6\textwidth]{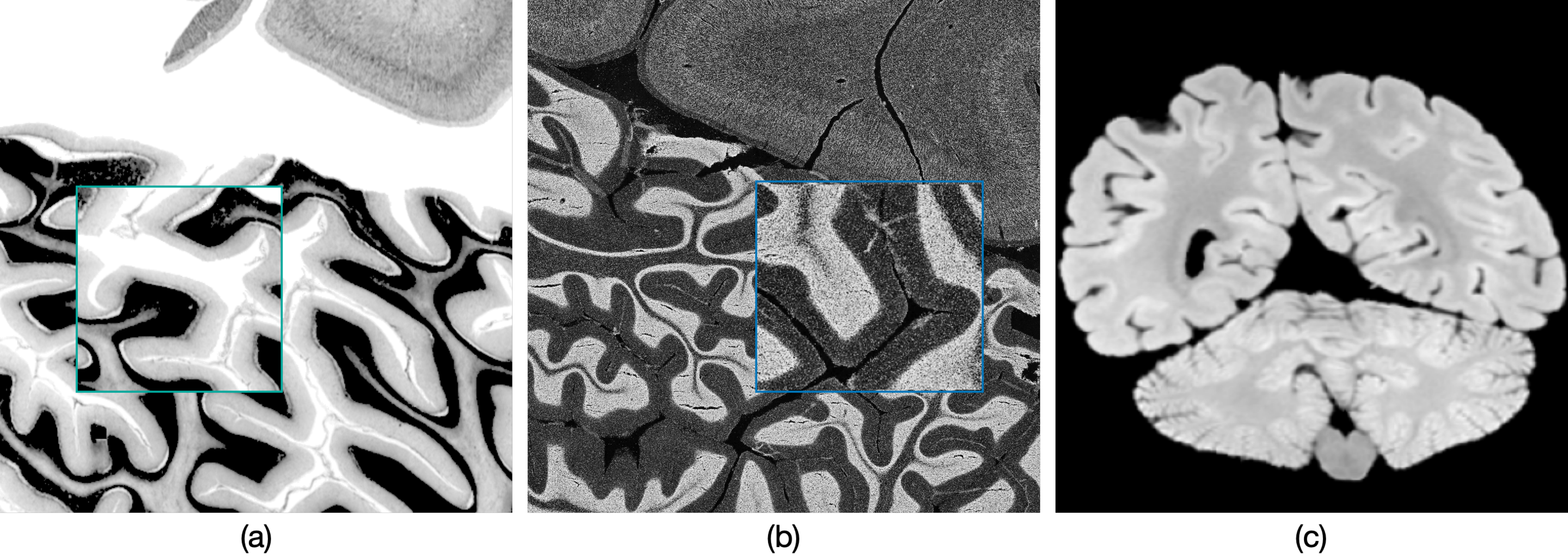} \\
\end{tabular}
\caption{Visual difference between 20-micron BigBrain (a), Allen Brain (b) histologies, and 1mm BigBrain MRI (c). Different staining methods, noise present in Allen Brain, and tissue tears in BigBrain cause large visual discrepancies between the two domains. Whereas with the coarser resolution MRI, it is impossible to observe the neural pathways.}
\label{fig:vis}
\vspace{-0.5cm}
\end{figure}

Recent advancements in convolutional neural networks (CNNs) have shown significant progress in computer-aided medical imaging analysis~\cite{schmahmann1999stereotaxic,schmahmann2000mriatlas,Sereno19538}. However, obtaining expert-level annotations for medical datasets is expensive and time-consuming, particularly for high-resolution histology data~\cite{mottolese2013mapping,schmahmann2019theory}, such as the BigBrain atlas~\cite{amunts2013bigbrain}, an open-source 3D human brain model for which a comprehensive cerebellum segmentation is lacking. Unlike the common 1mm MRI, BigBrain atlas is scanned at 20\micron to visualize the highly compact and convoluted cerebellar foliations (as shown in Figure~\ref{fig:vis} (a) and (c)). Expert manual annotation of the cortical structures is time-consuming~\cite{van2019strategies}.

To address this limitation, unsupervised domain adaptation (UDA) is a popular approach that transfers annotation knowledge from an existing labelled dataset (the source domain) to an unlabelled target domain~\cite{guan2021domain,ren2019unsupervised,fang2020dart,tellez2019quantifying}. In this work, we explore unsupervised semantic segmentation for the BigBrain (\BB) cerebellum, using the Allen Brain atlas~\cite{ding2016comprehensive} (\AB) as the source domain. However, this task presents additional challenges. First, each domain only contains cerebellum sections from a single subject, and these two subjects are demographically distinct. This heterogeneous nature results in large domain variance. Second, images in the two domains use different staining methods to visualize neurons, resulting in visual discrepancies. as indicated in Figure~\ref{fig:vis}: different color intensity distribution, the presence of noise in \AB, and damage in \BB. Semantic segmentation in medical imaging primarily relies on low-level features (texture, intensity, etc.)~\cite{guan2021domain}. Discrepancies in these low-level features bring extra challenges. Lastly, there are only 24 annotated cerebellum sections in \AB. Without abundant data, it is hard for the model to generalize well on the target dataset. 

To address these challenges, we propose BUDA-Seg, a Bayesian framework that quantifies the uncertainty in pseudo labels from a probabilistic perspective and guides the segmentation model to learn from regions with the most confidence. Concretely, our framework uses a pre-trained segmentation model on the \AB to generate pseudo labels for the \BB. To offset the noise and inconsistency in the pseudo labels, the framework outputs a variance map which can rectify the pixel-level noise in the pseudo labels. We then utilize variational sampling to approximate the true distribution of segmentation annotations for \BB. To our knowledge, our proposed framework is the first to achieve UDA for semantic segmentation with limited data in both source and target domains. Our contributions are summarized in the following~\footnote{Our code and data are available online at https://anonymous.4open.science/r/BUDA-Seg-8D75/README.md}:
\begin{itemize}
    \item We propose BUDA-Seg, a framework that uses Bayesian learning to improve UDA in semantic segmentation with limited data. It addresses the challenges posed by the noisy pseudo labels and distributional discrepancies between domains. 
    \item We demonstrate the effectiveness of our approach through quantitative and qualitative experiments on high-resolution cerebellum atlas and electron microscopy datasets. Our generated annotations for the cerebellum dataset will be publicly released for future research.
\end{itemize}

\section{Related Work}
\label{relatedwork}
Our research focuses on the task of unsupervised domain adaptation in semantic segmentation and specifically uses the label-explicit adaptation approach, which relies on the pseudo label.

\textbf{Label-explicit adaptation} uses the segmentation model trained on the source domain to generate pseudo labels for the target domain. Pseudo labels are further improved through iterative self-training. Hoffman et al.~\cite{hoffman2018cycada} adopt CycleGAN~\cite{zhu2017unpaired} to transfer the visual appearance of the source domain to the target domain, followed by iterative self-training to refine the segmentation outputs. Similarly, Yang et al.~\cite{yang2020fda} adopt Fourier transform on the two domains and then replace low-frequency components of the target domain with the source domain. The resulting target images have a visual appearance resembling the source domain. In~\cite{yang2021context,zou2018unsupervised,yan2019edge}, attention modules are used, where spatial and context information is used to locate and capture relevant semantics between two domains. Their works primarily concern UDA in natural scene images where the semantic information is on a high level (car, road, etc.). However, in histological data, the semantic information is at a low level (color, intensity, curvature, etc.), and using pseudo labels directly for training does not preserve the low-level features well, as we will show in our experimental results. Wu et al.~\cite{wu2021uncertainty} studied UDA in microscopy images by inserting dropout layers~\cite{gal2016dropout} to the UNet~\cite{ronneberger2015u} to generate pseudo labels with increased diversity. However, their approach requires multiple forward passes during both training and testing, imposing a heavy computational burden and limiting scalability to larger medical datasets. In comparison, our approach increases the diversity by sampling in the output predictions, requiring only one forward pass per data point and offering advantages in terms of run time and GPU consumption.

\begin{figure*}[ht!]
\centering
\setlength{\tabcolsep}{9pt} 
\begin{tabular}{cc}
\includegraphics[width=10cm]{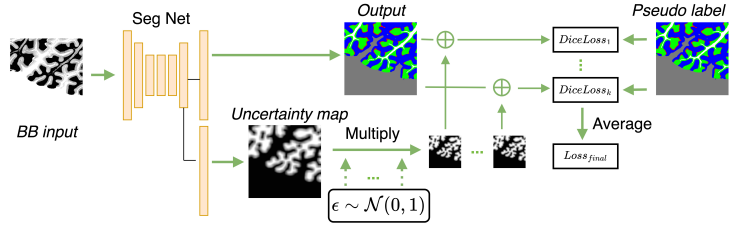} \\
\end{tabular}
\caption{Overview of our Bayesian uncertainty learning framework. For \BB input, Seg Net generates an output map, along with an uncertainty map. Repeated variational sampling is performed to adjust the uncertainty map and add it back to the output. Dice loss is calculated between each sampling and the pseudo label. The final loss is obtained by averaging over all Dice losses.}
\vspace{0.5cm}
\label{fig:training}
\vspace{-0.5cm}
\end{figure*}

\section{Method}
\label{method}
We begin by pre-training a segmentation network on \AB in a supervised manner. We then use the pre-trained network to generate pseudo labels for \BB. Since the domain gap between the two datasets is large and the amount of available data is limited, the resulting pseudo labels are error-prone. To address this issue, we propose a Bayesian uncertainty learning approach, as shown in Figure~\ref{fig:training}. Specifically, we have the $Seg Net$ output an uncertainty map to identify pixels that might have large losses between the noisy pseudo labels and predictions.

To implement this approach, we feed a \BB image $x \in \mathbb{R}^{W\times H}$ into $Seg_B$ to obtain an output prediction $O\in \mathbb{R}^{W\times H\times C}$ and an uncertainty map $V\in \mathbb{R}^{W\times H\times C}$, where $W, H$ are the spatial resolution and $C$ is the number of classes. We then perform a Gaussian sampling $\hat{O}_u \sim \mathcal{N}(O_u, V^2_u)$ for each pixel $u \in W\times H\times C$ to obtain an uncertainty-guided output. To ensure that gradients for both the output and uncertainty map could be back-propagated for training, We implement the sampling as follows:
\vspace{-0.2cm}
\begin{equation}\label{eqn:sampling}
\hat{O}_{k,u} = O_u + \epsilon _k V_u \ ,
\end{equation}
\vspace{-0.1cm}
where $\epsilon _k \sim \mathcal{N}(0,1)$, and $k$ is the $k$-th sampling. This stochastic sampling does not require extra forward passes to the neural net, and it can be implemented in parallel on the GPU for all pixels and all images in a batch, making it computationally efficient and using only a fraction of computing resource. We then calculate the probability $P_{k,u}$ using softmax of $\hat{O}_{k,u}$, and compute the Dice loss:
\vspace{-0.2cm}
\begin{equation}\label{eqn:dice}
Dice\ Loss_{k} = 1 - \frac{2\sum_{u\in W\times H}P_{k,u}\hat{Y} _u}{\sum_{u\in W\times H}P^2_{k,u} + \sum_{u\in W\times H}\hat{Y}^2 _u} \ ,
\end{equation}
\vspace{-0.1cm}
where $\hat{Y} _u$ is the pseudo label at pixel location $u$. This sampling is repeated for a total of $K$ times. The final loss is averaged over all samplings: 
\vspace{-0.2cm}
\begin{equation}\label{eqn:final_loss}
Loss_{final} = \frac{1}{K}\sum_{k=1}^K Dice\ Loss_k \ .
\end{equation}
\vspace{-0.1cm}
To leverage iterative self-training, we use the current predictions of \BB as the new pseudo labels for the next training iteration. We repeat a total of $T$ iterations. 

\textbf{Discussion:} The rationale for our proposed uncertainty learning approach is that pseudo labels for \BB are error-prone, and forcing the model to learn from these labels would aggravate the noise. In contrast, the introduced uncertainty map allows the model to rectify pixels with high segmentation loss that might be due to the noisy pseudo label, and thus assigns higher variance to offset the segmentation loss. Note that, under these conditions, variance is analogous to the absolute value of uncertainty. We therefore use the two terms interchangeably. Our segmentation network is modelled in a Bayesian way, in order to provide calibrated uncertainty of the pseudo labels, hence named BUDA-Seg.

\begin{figure}[t!]
\centering
\setlength{\tabcolsep}{9pt} 
\begin{tabular}{cc}
\includegraphics[width=.6\textwidth]{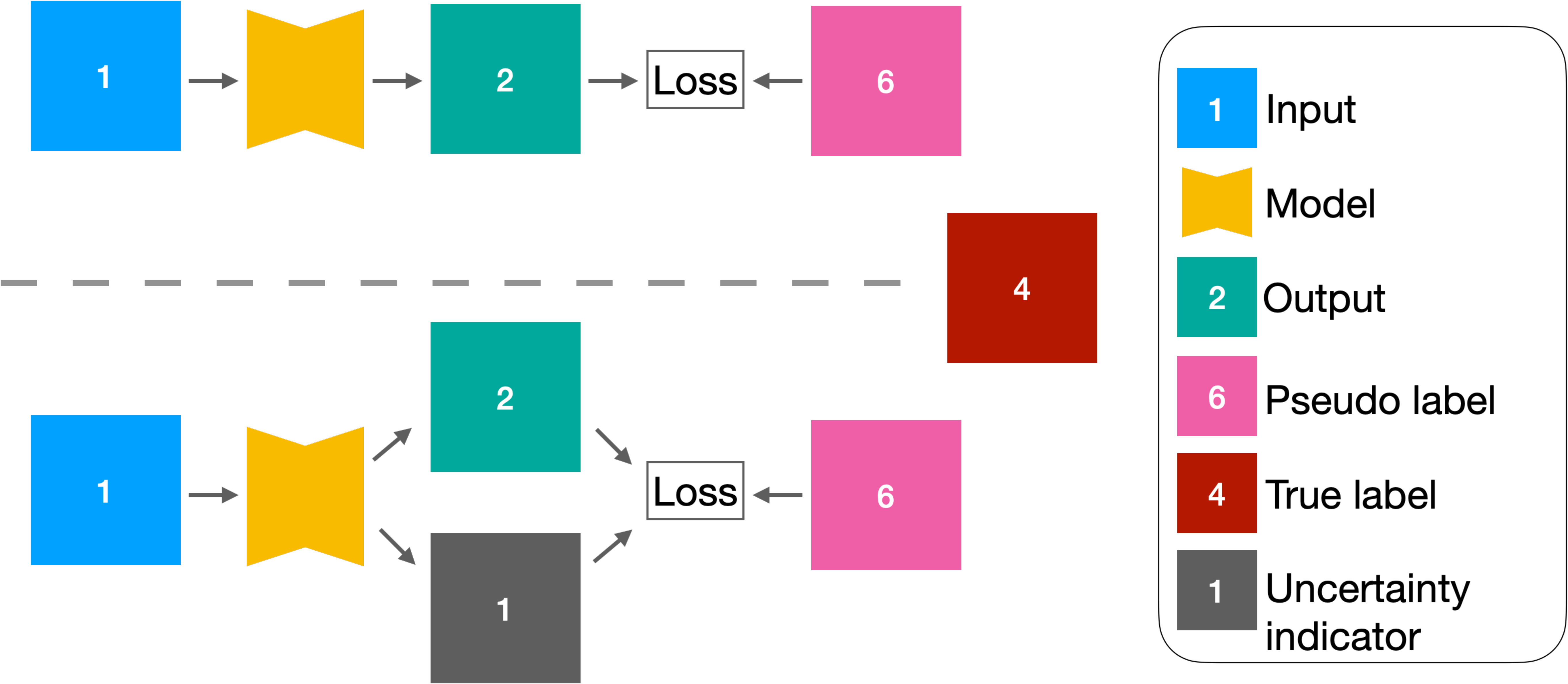} \\
\end{tabular}
\caption{An intuitive example of regular training (top row) and uncertainty learning (bottom row). The uncertainty indicator can offset a large loss between output and noisy pseudo label}
\vspace{0.5cm}
\label{fig:example}
\vspace{-0.8cm}
\end{figure}

To provide an intuitive explanation of our Bayesian uncertainty training, We provide a simple example in Figure~\ref{fig:example}. Suppose we have a target input $x = 1$, feeding into a model $f(\cdot)$ to obtain an output $o = 2$. Considering we have a noisy pseudo label $\hat{y} = 6$, whereas the unknown ground truth $y = 4$. Forcing the model to minimize the \textit{false} distance between $\hat{y}$ and $o$ would not help training converge. Therefore, we propose an uncertainty indicator $v$ generated by $f(\cdot)$ to offset the large loss. For a concrete example, $v = 1$, and we have the new output $o' = 3$. The reduced loss $3=6-(2+1)$ could help $f(\cdot)$ generate output closer to the ground truth and thus stabilize the training.

\section{Experiments}
\label{experiments}
In this section, we describe the datasets used for our experiments. We include the setup for training, metrics used for evaluation, and choices of baseline approaches in the Appendix.

\textbf{Datasets} The Allen Brain (\textit{AB}) adult human brain atlas used here, consists of 668 Nissl-stained coronal sections of the left hemisphere at 200\micron spacing, scanned at $1 \times 1$\micron per pixel, with 106 annotated cerebral slices (specimen 708084, 34 y.o. female, Modified Brodmann labels)~\cite{ding2016comprehensive}. We downloaded 24 slices containing the cerebellum and created patches of $1000\times 1000$ pixel resolution. A total of 1006 patches are created to form the source domain data. We release these data online.

The BigBrain (\textit{BB}) is an open-source 3D model of a whole human brain at 20\micron isotropic resolution, reconstructed from high-resolution scans of 7,404 histological sections marked for cell bodies with Merker (silver) stain (65 y.o. male)~\cite{amunts2013bigbrain}. Each section is about $6500\times5700$ pixel resolution. The differences in staining (contrast) and in the original resolution of the images add complexity to the model. Given the high visual similarity between adjacent 20\micron-thick slices containing the cerebellum, we kept 22 sections with large differences and created patches of $1000\times 1000$ pixel resolution. A total of 600 patches are created to form the target domain data. Expert annotation on 3 independent sections are split into 75 patches, which are used for evaluation only.

% \vspace{-0.5cm}

\begin{figure*}[ht!]
\centering
\setlength{\tabcolsep}{10pt} 
\begin{tabular}{c}
\includegraphics[width=.8\textwidth]{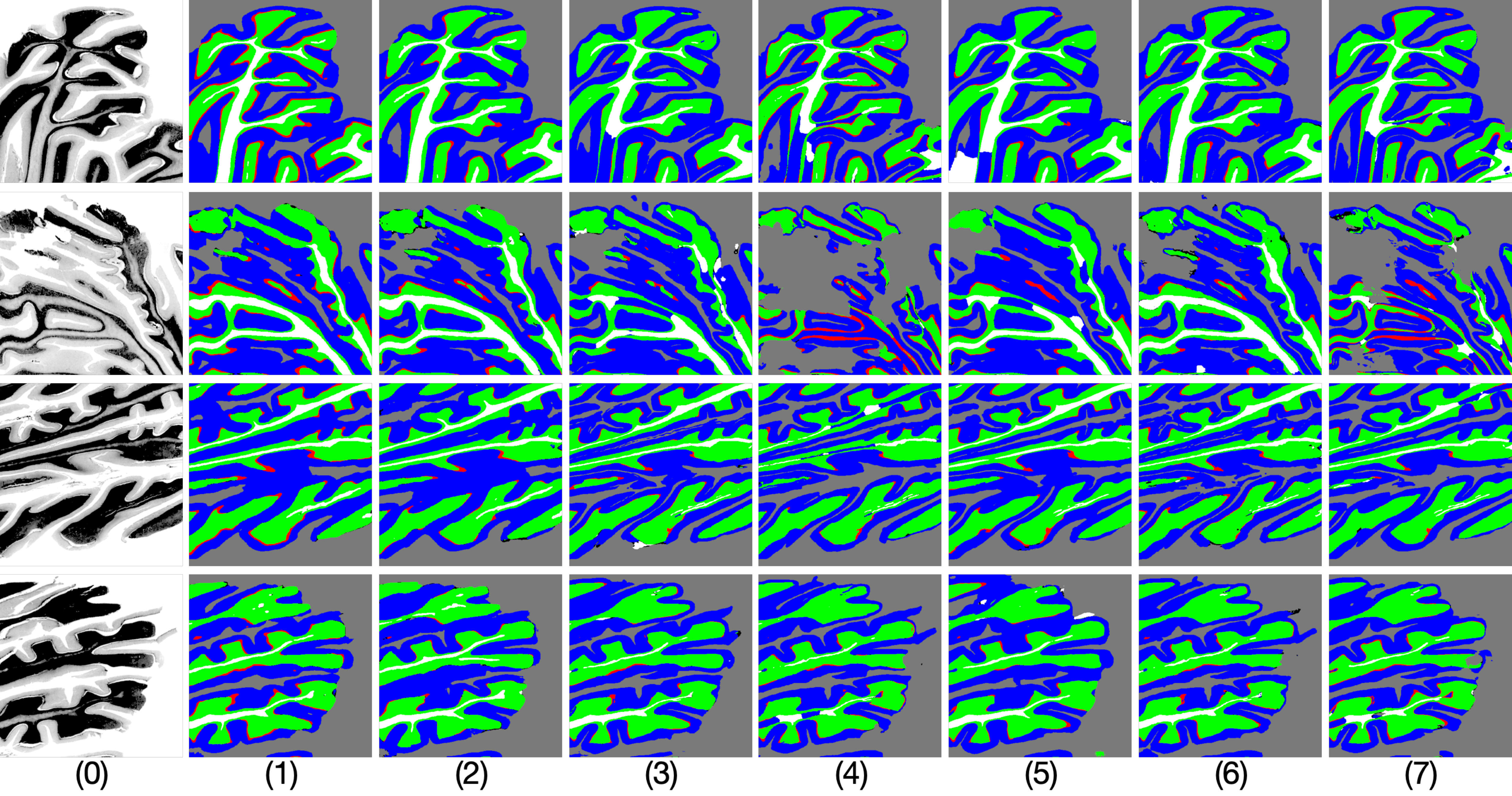} \\ 
\end{tabular}
\caption{Segmentation results comparison. (0) original \BB inputs; (1) our approach; (2) Wu et al.~\cite{wu2021uncertainty}, (3) Yang et al.~\cite{yang2021context}, (4) Hoffman et al.~\cite{hoffman2018cycada}, (5) Chung et al.~\cite{chung2022maximizing}, (6) Kamnitsas et al.~\cite{kamnitsas2017unsupervised}, (7) Zou et al.~\cite{zou2018unsupervised}. Gray = background; blue = molecular layer; red = Purkinje cell layer; green = granular layer; white = white matter. Additional results are in Appendix and code repo.}
\label{fig:result}
\vspace{-0.5cm}
\end{figure*} 

% \vspace{-0.5cm}
\section{Results}
\label{sec:results}
\vspace{-0.2cm}
In this section, we present our main segmentation results and compare them with other baseline approaches. We also report an ablation study, a qualitative assessment, and an extensive experiment on another medical dataset to demonstrate the effectiveness of our framework.

\vspace{-0.3cm}
\begin{table*}[ht!]
 \centering
 \caption{Quantitative comparisons. Mean Dice loss, and mean IoU, with standard deviations (\%). Our proposed BUDA-Seg achieves the lowest Dice loss and highest IoU.}
 \label{tab:mean-dice-loss}
 \setlength{\tabcolsep}{4pt} 
 \renewcommand{\arraystretch}{1}
% \begin{small}
\fontsize{8.5}{8}\selectfont
\begin{tabular}{l|cc}
 \toprule
  & Dice loss $\downarrow$ & IoU $\uparrow$ \\
 \midrule
BUDA-Seg (Ours) & \textbf{55.7 \ppm 0.7} & \textbf{73.22 \ppm 1.0} \\

Wu et al.~\cite{wu2021uncertainty} & 56.8 \ppm 1.1 & 70.54 \ppm 1.6 \\

Yang et al.~\cite{yang2020fda} & 57.9 \ppm 1.3 & 67.42 \ppm 1.9 \\

Hoffman et al.~\cite{hoffman2018cycada} & 60.2 \ppm 2.0 & 67.16 \ppm 2.5 \\

Chung et al.~\cite{chung2022maximizing} & 57.7 \ppm 1.4 & 65.9 \ppm 2.1 \\

Kamnistas et al.~\cite{kamnitsas2017unsupervised} & 61.3 \ppm 2.2 & 65.9 \ppm 2.4 \\

Zou et al.~\cite{zou2018unsupervised} & 60.7 \ppm 1.5 & 67.1 \ppm 1.9 \\
 \bottomrule
\end{tabular}
% \end{small}
\vspace{-0.5cm}
\end{table*}

\subsection{Main Results}
We display our segmentation results on \BB in Figure~\ref{fig:result}. BUDA-Seg (column (1)) provides the most stable predictions compared to other baselines. For example, in row 2, the input \BB patch has a large portion of tears and artifacts caused by sectioning. BUDA-Seg shows adaptation to these regions while still producing a meaningful semantic segmentation. When data have elongated structures (long stripes such as row 1), BUDA-Seg provides consistent segmentation results, whereas other approaches generate discontinuous predictions. Our approach also faithfully preserves the Purkinje cell layer (indicated in red), which is a thin and sparse layer between the granular layer (green) and the molecular layer (blue). Other approaches either miss or incorrectly place the Purkinje cell layer. The cerebellum has a highly convoluted structure, and being able to distinguish the boundaries is crucial for understanding cerebellar structure and function. BUDA-Seg achieves overall better fine-grained segmentation results by preserving the space and gaps and thus having less connected regions, such as the space between curvatures in rows 3 and 4. We further provide the mean Dice loss and IoU in Table~\ref{tab:mean-dice-loss}. BUDA-Seg achieves the lowest loss by 2.0\% loss reduction compared with Wu et al.~\cite{wu2021uncertainty}. Our approach also obtains the highest IoU with more than 3.7\% improvement. In addition, our approach requires one forward pass to quantify the uncertainty, thus resulting in over 60\% GPU memory efficiency and over 50\% time efficiency compared with Wu et al.~\cite{wu2021uncertainty}, as is shown in Table~\ref{tab:speed_comparison}.

\vspace{-0.3cm}
\begin{table}[ht!]
 \centering
 \caption{GPU usage in gigabytes (GB) and training speed in seconds (s) comparison on a batch size of 1 image with $K=4$. Our approach is over 60\% memory efficient, and 50\% training time speedup.}
 \label{tab:speed_comparison}
 \setlength{\tabcolsep}{6pt} 
 \renewcommand{\arraystretch}{1}
\begin{small}
\begin{tabular}{lcc}
 \toprule
 \bf Method & \bf GPU usage(GB)$\downarrow$ & \bf Training speed(s)$\downarrow$ \\
 \midrule
 BUDA-Seg & 3.06 & 0.371 \\
Wu et al. & 8.01 & 0.708 \\
 \bottomrule
\end{tabular}
\end{small}
\vspace{-1.0cm}
\end{table}

\begin{figure}[ht!]
\centering
\setlength{\tabcolsep}{9pt} 

\begin{minipage}[b]{0.44\textwidth}
    \includegraphics[width=\textwidth]{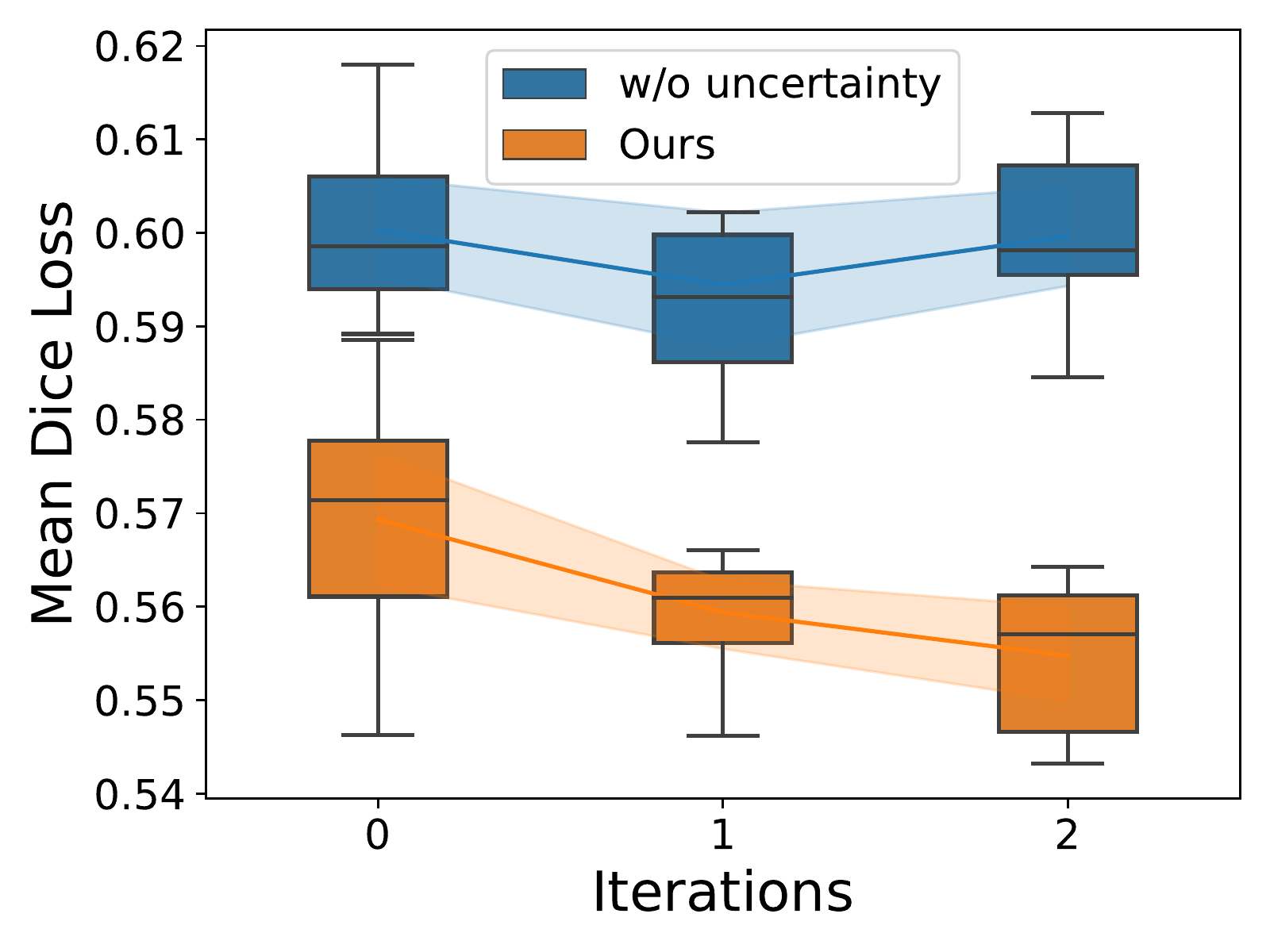}
    \caption{Boxplot of mean Dice loss comparison between our approach and without uncertainty learning at each iteration.}
    \label{fig:dice_loss}
  \end{minipage}
  \hfill
  \begin{minipage}[b]{0.46\textwidth}
    \includegraphics[width=\textwidth]{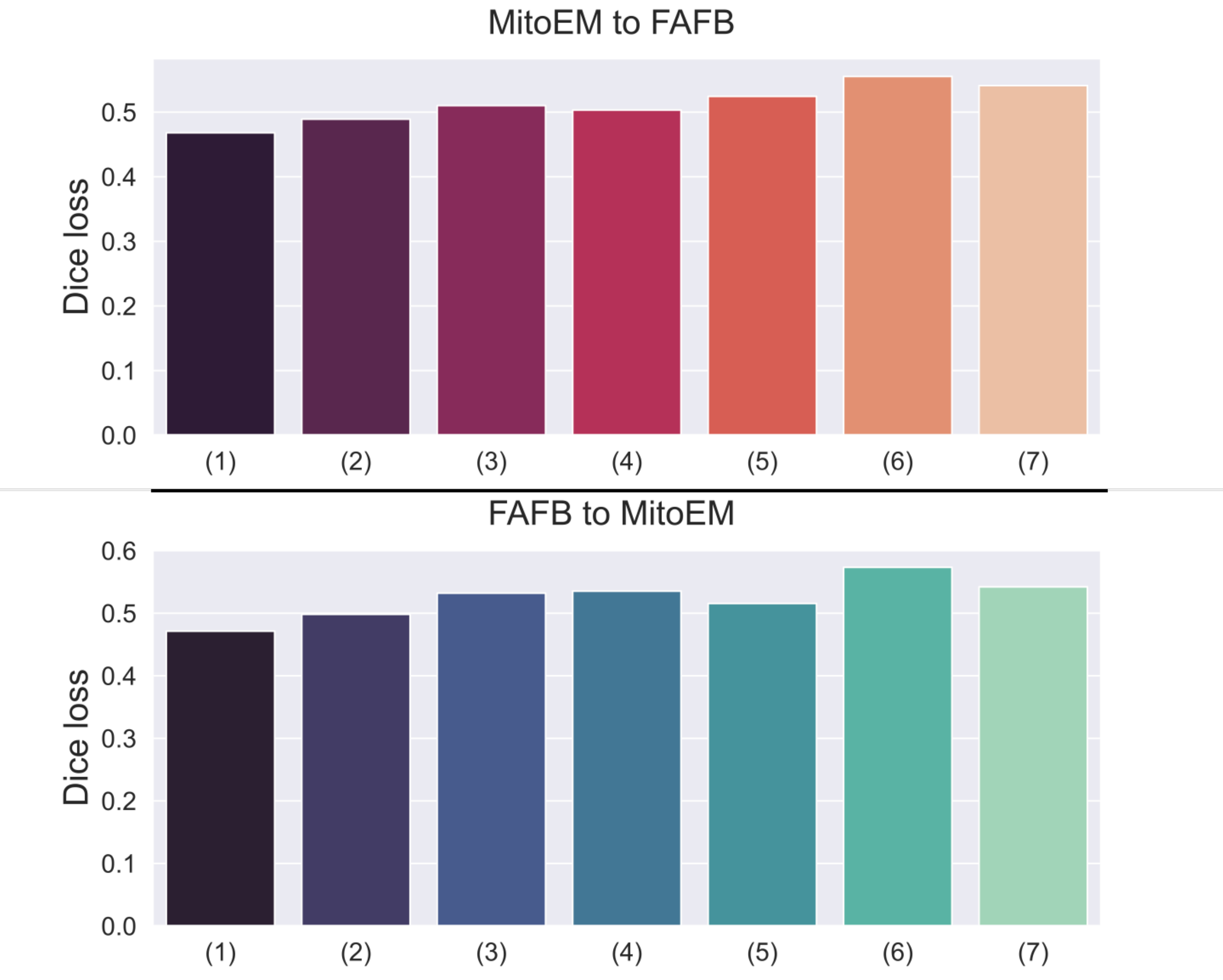}
    \caption{Dice loss comparison when transferring from MitoEM~\cite{wei2020mitoem}(source domain) to the FAFB~\cite{zheng2018complete}(target domain) in the upper figure, and vice versa in the lower figure.}
    \label{fig:qd}
  \end{minipage}
\vspace{-0.3cm}
\end{figure}

\subsection{Effectiveness of Bayesian learning}
To validate the effectiveness of our proposed Bayesian learning strategy, we compute the Dice loss at each training iteration and compare our approach against a baseline approach without uncertainty learning. As shown in Figure~\ref{fig:dice_loss}, in the first iteration, our approach effectively mitigates regions with large segmentation loss and trains the model to offset the inconsistency between the input data and pseudo label with an uncertainty map. In contrast, the baseline approach does not offer a flexible way to rectify the inconsistency, resulting in a larger Dice loss. 

Furthermore, with iterative training, our approach enables models to learn from uncertainty, resulting in a steadily decreasing Dice loss after each iteration. In contrast, the baseline approach cannot adapt the model to steadily learn between the input and pseudo label after each iteration, causing an increased loss after the third iteration. These results demonstrate the effectiveness of our proposed Bayesian learning strategy in improving segmentation performance and mitigating inconsistency.

\vspace{-0.1cm}
\subsection{Qualitative Visual Assessment} 
\vspace{-0.1cm}
We randomly sampled 20 \BB input images with corresponding segmentation results from 4 competitive approaches (BUDA-Seg, \cite{wu2021uncertainty}, \cite{yang2021context}, \cite{chung2022maximizing}). An independent expert rater assigned a ranking score of 1 to 4 to each segmentation result based on its quality, with 4 indicating the highest quality and 1 the lowest. Our approach received the highest overall score of 63, while the other approaches obtained scores of 60, 43, and 34, respectively. These results demonstrate the superior performance of our approach in generating high-quality semantic masks. 

\vspace{-0.1cm}
\subsection{Extensive Experiments}
\vspace{-0.1cm}
To further validate the effectiveness of our proposed framework on different datasets, we apply BUDA-Seg to two publically available datasets MitoEM~\cite{wei2020mitoem} and FAFB~\cite{zheng2018complete}. The MitoEM~\cite{wei2020mitoem} dataset consists of 500 sections of a rat brain, with each section having a resolution of 4096$\times$4096. The FAFB~\cite{zheng2018complete} dataset is a serial section of a fly brain scanned at 40$nm$, and we use 500 cropped sections with a resolution of 4000$\times$4000. We conduct two experiments: one where we transfer the annotation from the source domain MitoEM~\cite{wei2020mitoem} to the target domain FAFB~\cite{zheng2018complete} (shown in the upper plot of Figure~\ref{fig:qd}), and another where we transfer the annotation from FAFB~\cite{zheng2018complete} to MitoEM~\cite{wei2020mitoem} (shown in the bottom plot of Figure~\ref{fig:qd}). In both experiments, BUDA-Seg achieves the lowest Dice loss, demonstrating its effectiveness and generalizability across different datasets.

\vspace{-0.1cm}
\section{Conclusion}
\vspace{-0.2cm}
This paper introduces BUDA-Seg, a Bayesian learning framework for UDA in semantic segmentation. Our approach addresses the challenge of distributional discrepancy between the source and target domains by estimating prediction uncertainty and adapting the training process from noisy pseudo labels. We validate our method on a high-resolution cerebellum atlas dataset and two electron microscopy images, demonstrating its effectiveness in achieving state-of-the-art segmentation results in UDA.

% estimates the prediction uncertainty, and rectifies the training from the noisy pseudo labels when the source domain and target domain have a large distributional discrepancy. We demonstrate the effectiveness of our approach on a high-resolution cerebellum atlas dataset, as well as two electron microscopy images. 
%In the future, we will explore how BUDA-Seg can be extended to sub-cortical cerebellar structures, as well as generalized to other regions of the brain to produce an atlas with more consistent labels.  

\bibliographystyle{ieeetr}
\bibliography{mybibliography}
\end{document}